# Flaws in Flawlessness: Perfectionism as a New Technology Driven Mental Disorder

*Completed Research Paper*

**Darshana Sedera**
Southern Cross University
Gold Coast, Australia
darshana.sedera@gmail.com

**Sachithra Lokuge**
RMIT University
Melbourne, Australia
ksplokuge@gmail.com

## Abstract

*Today, technologies, devices and systems play a major role in our lives. Anecdotal commentary suggests that such technologies and our interactions with them create a false sense of perfectionism about life, events and its outcomes. While it is admirable to strive for better outcomes; constant and sometimes unrealistic expectations create a psychological condition commonly known as 'Perfectionism'– the fear of not doing something right or the fear of not being good enough. In this paper, based on the Diagnostic Statistical Manual of Mental Disorders (DSM-III), we conceptualize 'digital perfectionism' as an emerging disorder, that is specific to our increasing interactions with tools and technologies. By using a sample of 336 individuals, this study makes valuable early insights on digital perfectionism, its conceptualization and its effects on the individuals.*

## Introduction

Since the dawn of civilization, individuals are relentlessly attempting to improve themselves and how others perceive them (Chua and Chang 2016; Salleh et al. 2009). In such endeavors, information technologies (IT) play a major role in mimicking nearly all daily activities of one's life, ranging from personal events to professional activities (Carter and Grover 2015; Chua and Chang 2016). Therein, the information systems (IS) scholars highlight the role of technologies in providing individuals with a plethora of technology options to increase their productivity, efficiency, minimize errors, increase quality of outcomes, manage and track their engagements (Yoo 2010). Over the past decade, such technologies that were predominantly focused on organizational interventions, have begun to impact individual behaviors (Carter and Grover 2015). As such, more and more applications that are embedded in laptops, smart mobile phones, smart devices, digital cameras and social media platforms encourage 'corrective' or 'recommended' behaviors to enhance or rectify the 'original' state of behaviors or outcomes of the individual. Examples of such actions include auto-correcting photographs, suggesting exercises, comparative behaviors with online social groups and application of electronic work templates that enhance the presentation quality.

While it is noble to seek perfectionism, such frequent engagements of corrective (or optimal) behavior through constant digital interventions create a pretense of perfectionism. For example, a consumer survey identified that an individual on average takes up to eleven pictures before uploading the 'perfect selfie' on their social media platform (FHE Health 2017). In an extreme example, a selfie obsessed teenager who took 200 selfies committed suicide after failing to capture 'the perfect selfie' (Molloy 2014). In 2019, BBC Futures represented studies of Sedera and Lokuge (2018) and Oakes (2019), highlighting the necessity of researching psychological disorders arising from digitalization – especially within the younger adults, where the social pressures are further exerted by the use of such technologies (Sweeten et al. 2018).

Amongst the teens (though not limited to), the heightened engagement with social media and digital technologies have created a false sense of perfectionism about life events (Griffiths and Balakrishnan 2018), making it difficult for individuals to deal with the realities of failure (Levine et al. 2017; Molloy 2014).

Such behaviors that are entangled with technologies have led to a society that is less tolerant of errors or imperfections, increasing the fears of imperfections amongst individuals – a condition clinically known as perfectionism (also commonly known as *Atelophobia*). This is recognized as a mental disorder in the 3rd edition of the diagnostic statistical manual of Mental Disorders (DSM-III) (American Psychiatric Association 2013). In recent times, social scientists warn that, with the heightened digital engagements, a mirage of perfectionism is on the rise (Ellison et al. 2006; Jain and Mavani 2017), creating illnesses like depression, loneliness and anxiety (Griffiths and Balakrishnan 2018). Moreover, a heightened focus on perfectionism shifts the attention from quality of the content to the quality of a superficial endorsement of the *presentation*. As such, studies highlight that presentation of digital content receives far greater emphasis compared to the originality of the idea (Chua and Chang 2016). Similarly, there is a growing consensus that individuals are less willing to take risks with the fear of making mistakes (Levine et al. 2017).

While there is anecdotal commentary on the detrimental effects of perfectionism in everyday practices of most individuals, to the best of our knowledge, there has been no academic studies in IS outlets on the notions of perfectionism in the digital world. In this research, we define perfectionism behavior of the individual observed when dealing with digital content as *digital perfectionism* – where the word 'digital' highlights the context that it is rooted. In doing so, we draw parallels with the traditional psychological illness of perfectionism, established in the DSM-III (American Psychiatric Association 2013). This paper investigates the notion of digital perfectionism by developing measures consistent with the established guidelines of the DSM-III (American Psychiatric Association 2013), to assess whether individuals demonstrate characteristics of perfectionism in producing digital content. In doing so, we draw insights from the fundamental studies on perfectionism in psychology (e.g., Frost et al. 1990; Pacht 1984) and sociology (e.g., Greenspon 2000). We highlight that the DSM instruments are regularly employed in IS behavioral disorder studies of internet addiction (Kloker 2020), smart phone addition (Kuem et al. 2020) and in general to understand the 'dark-side of IS' (D'Arcy et al. 2014).

The paper proceeds in the following manner. First, the study provides the theoretical foundation and derives the a-priori model of the study. Next, the paper introduces the data collection and the sample employed to test the a-priori model. Subsequently, the data analysis is presented. A post-hoc multigroup analysis is also carried out to highlight further insights of the study. The results of the study are described next, drawing conclusions for research and practice, finally, summarizing the limitations of the study.

## Theoretical Basis of Digital Perfectionism

Frost et al. (1990) define perfectionism as the setting of excessively disproportionate standards for performance accompanied by overly critical self-evaluations. Past studies of perfectionism highlight different types of perfectionism behaviors such as positive vs. negative (Terry-Short et al. 1995), or healthy vs. unhealthy (Stumpf and Parker 2000) perfectionism behaviors. The characteristics of perfectionism has been described as excessive orderliness, disproportionate standards and an inability to feel a sense of satisfaction. As per Hamachek (1978, p. 27) positive perfectionists set disproportionate standards for themselves, yet, adjust them as the situation permits. However, negative perfectionists never feel that anything is done completely enough or well enough (Frost et al. 1990).

The early conceptualizations suggest perfectionism to be a unidimensional concept (Burns 1980). However, recent studies have identified that perfectionism is multidimensional in nature (Frost et al. 1990; Hewitt and Flett 1991). Frost et al. (1990) define perfectionism using a multi-dimensional model known as the 'Frost's multidimensional perfectionism scale' (FMPS). Even though FMPS is considered as an established measure of perfectionism, there is much debate amongst researchers regarding the optimum number of dimensions that are required to measure perfectionism. Studies such as Parker and Adkins (1995) and Rhéaume et al. (1995) argue for using the original six factor model for measuring perfectionism. A number of subsequent studies have confirmed that, the *six dimensions can be consolidated into four* mutually exclusive constructs with adequate discriminant and convergent validity (Stöber 1998; Stumpf and Parker 2000). The following parsimonious four constructs are employed in this study: (i) Mistake uneasiness, (ii) Disproportionate standards, (iii) Criticism antipathy and (iv) Excessive orderliness (e.g., Stöber 1998;

Stumpf and Parker 2000). The studies that employ only four dimensions of FMPS have displayed a pattern of substantial correlations with related variables demonstrating the usefulness and the need of using only four dimensions of FMPS to measure perfectionism (Stöber 1998).

While anecdotal evidence of the existence and prevalence of the perfectionism behavior is aptly visible in relation to digital engagements, a scientific investigation based on the premise of 'traditional' perfectionism to explore 'digital' perfectionism requires logical reasoning. In order to extend the theoretical notions of traditional perfectionism behaviors (e.g., FMPS) and to distil and develop the differences of traditional perfectionism and digital perfectionism, we developed a framework that examines the tripartite relationship among the creator-content-audience. The framework presented in Table 1 therefore, allows us to seek differences between *digital* perfectionism vs. *traditional* perfectionism. Therein, we demonstrate an iterative relationship between the three fundamental entities through the use of digital tools.

**Table 1. Digital Perfectionism Considerations**

| | Entity | Considerations for the 'digital' context |
|---|---|---|
| 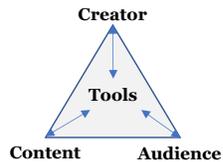 | Creator | Technology as the creator, augmented capabilities of the creator, co-creation, creator communities |
| | Content | High rate of content creation, content creation platforms, automated content creation |
| | Audience | Large numbers, unintended audiences, anonymity of the audience |
| | Tool | Creation tools, collaboration tools, dissemination tools, assessment tools |

**Table 1. Digital Perfectionism Considerations**

As per Table 1, digital perfectionism occurs with the interaction of the three aspects: (i) *creator* – the individual who allegedly displays the notions of perfectionism, (ii) *content* – the entities that are being adjudicated and (iii) the *audience* – those who adjudicate the contents produced by the individual. While these three notions are conceptually common to both traditional and digital context, the perfectionism behavior in the digital context is augmented through the example considerations of Table 1. In the digital context (as compared to traditional), (i) the creator, potentially embedded in a community of creators or technology assuming the role of the creator is provided with applications, systems, features, tools and techniques not only to engage in rectifying behaviors, but also to reach to audiences and to disseminate content. The (ii) the rate of such content creation is high in digital, (iii) the increased digital connectivity through digital platforms create a larger social circles compared to the traditional context (Sedera et al. 2016), and (iv) the quality of the individual's work may be adjudicated against material produced by those who are proficient. Such deviations in the digital context further inspired us to investigate perfectionism behaviors in the digital context.

## Deriving the A-priori Research Model

The a-priori research model of digital perfectionism is conceptualized using the foundational works of Frost et al. (1990), conceiving digital perfectionism as a second-order construct that consists of four dimensions. The four dimensions of the a-priori model are entrenched in the aforementioned perspectives in Table 1 of the digital context. Therein, we argue for the iterative relationships between the creator-content-audience, where the creator would feel a heightened state of the following dimensions. Figure 1 depicts the digital perfectionism, an a-priori research model, which consist of the four sub-constructs that are conceived and measured as a formative composite construct.[1]

### Mistake Uneasiness

This is a major characteristic that distinguish normal vs. neurotic perfectionists. While a normal perfectionist has wider latitude in allowing mistakes, the neurotic perfectionists are overly concerned with mistakes that even minor ones are likely to result in the perception that their standards 'have not been met.' The over concern for mistakes, according to Hamachek (1978), leads perfectionists to strive for their goals by a fear of failure, rather than a need for achievement. Burns (1980) also emphasize the importance of fear of mistakes in defining perfectionism and characterizes it as part of the dichotomous thinking style of depressives as performance must be perfect, or it is worthless and that any minor flaw constitutes failure. Furthermore, the sense of doubt has been extensively described in the literature on obsessional experiences as a feeling of uncertainty regarding an action or belief. Reed (1985, p. 115) classifies perfectionism with other characteristics of obsessive-compulsives which reflect a "reluctance to complete a task." The literature on digital literacy, digital creativity and digital media highlight that there is an increased level of overly critical evaluative tendencies amongst the individuals engaged in such digital spaces. For example, Padoa et al. (2018) highlight that individuals who recently became mothers attempt to promote the most positive representations of themselves in digital media perhaps to avoid being perceived as a "bad mother."

---

[1] Benitez et al., (2017, p. 2) demonstrate factor and composite measurement models that PLS path models contain. "*Factor models use reflective constructs and assume that the variance of a set of indicators can be perfectly explained by the existence of one unobserved variable and individual random error.*" Benitez et al., (2017) state that composite models are formed as linear combinations of respective indicators and they serve as proxy for the concept under investigation that is composed of a mix of indicators.

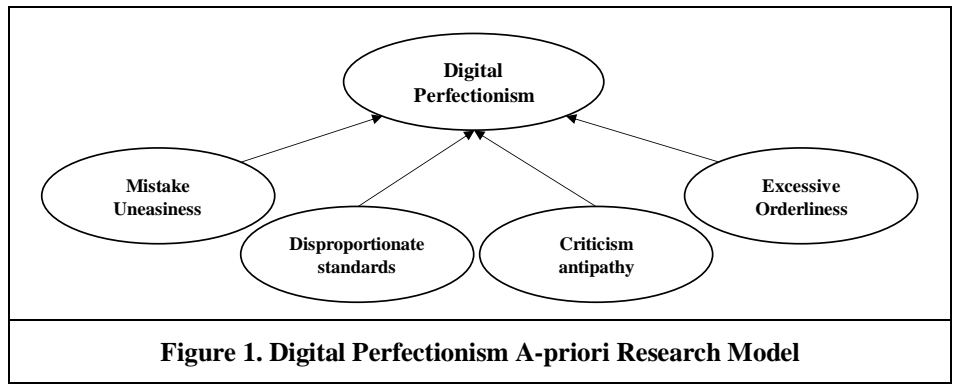

**Figure 1. Digital Perfectionism A-priori Research Model**

## Disproportionate Standards

Adherence to excessively high and disproportionate personal standards of performance is one of the most commonly discussed topics in perfectionism literature (Burns 1980; Pacht 1984). When asserting this, one must consider that setting and striving for high standards is certainly not extreme. In fact, such high standards may reflect a positive outlook on life (Pacht 1984). As per Hamachek (1978, p. 27), normal perfectionists are those who set high standards for themselves yet "feel free to be less precise as the situation permits." Neurotic or negative perfectionists set high standards but allow little leeway for making mistakes. The implication of this distinction is that perfectionism involves high standards of performance which are accompanied by tendencies for overly critical evaluations of one's own behavior. Recent studies highlight that individuals set highly unrealistic standards in relation to digital content (Lim 2016; Molloy 2014). Moreover, in relation to producing digital content, it is easy to enjoy a false sense of 'purity,' assisted by the features and functions of technologies and tools. Such tools, commonly available in technologies and social media platforms, provide a false sense of a 'perfect world' (Boyd 2007; Lim 2016) and creating disproportioned standards.

## Criticism Antipathy

Criticism antipathy relates to several theories of how individuals place considerable value on the views of their social circles (e.g., social influence theory). Perfectionism literature highlight that individuals tend be conditional to love and approval of their social circles (Hamachek 1978; Hollander 1965). Such studies argue that, to feel loved and supported, the individual attempts to perform at high levels of perfection and is not willing to accept criticism. Such situations are aggravated in the context of digitally connected societies, where there is an ever-growing level of social influence on individuals (Sedera et al. 2017a). For examples, studies employing the social influence theory (Kelman 1958) and social impact theory (Latané 1981) highlight that the (i) the number of individuals in the digital social network, (ii) perceived strength of the relationships and (iii) perceived immediacy of the social network have a strong impact on the individual's social expectations. Moreover, the alleged 'free-speech' in digital media and the intense intolerance in digital platforms have the potential to exacerbate this situation (Chua and Chang 2016).

### Excessive Orderliness

As per Hollander (1965, p. 96) excessive orderliness is a tendency to be "fussy and exacting" with an overemphasis on orderliness. Excessive orderliness is related to how the individual goes about the day-to-day task of meeting those standards and therefore, it is considered an important dimension of perfectionism. In this perspective, digital content receives special attention, especially when there is a greater possibility for an individual to organize (and re-organize) digital content. Sedera and Lokuge (2018) introduced the notion of 'digital hoarding' as a possible disorder, partially associated with the way individuals accumulate, store and dispose (or its reluctance to) digital content. Therein too, they highlighted the importance of keeping digital content 'organized.'

## Data Collection and Sample

The survey instrument included 21 reflective items (See Appendix A) arranged under the four dimensions of digital perfectionism, following the FMPS instrument and its extensions (Frost et al. 1990; Stöber 1998; Stumpf and Parker 2000). The complete instrument (without explanatory notes and instructions) is available in Appendix A. The instrument also captured respondent's: (i) age, (ii) gender, (iii) number of friends in social media and (iv) the level of digital engagement. In addition, 9 items were added to measure the dependent variable 'anxiety,' using the well-established Lovibond and Lovibond (1995) scale of depression, anxiety and stress scale (DASS). Further 4 items were included as 'global measures' for digital perfectionism, necessary for formative construct validation and model testing (Gable and Sedera 2009). All items were measured in a seven-point Likert scale with the end values ranging from 1 to 7. The instrument was pilot tested with a sample of 33 respondents. The pilot survey analysis resulted in addition of explanatory statements. For example, concerns were raised about the definition of digital content. As such, a new statement was added instructing the respondents as follows: "This survey assesses the digital content creations or engagements of yours." The instrument was circulated amongst a representative sample of the population of 800 individuals, yielding responses from 336 respondents (with a response rate of 24%). The demographics of the respondent sample are summarized in Table 2.

| Table 2. Demographics of the Respondent Sample | | | | | |
|---|---|---|---|---|---|
| **Gender** | **#** | **%** | **# of Associates** | **#** | **%** |
| Male | 148 | 44% | less than 20 | 40 | 12% |
| Female | 188 | 56% | 20-50 | 37 | 11% |
| **Content in context** | **#** | **%** | 51-100 | 64 | 19% |
| Mainly professional | 208 | 62% | 101-200 | 81 | 24% |
| Mainly personal | 128 | 38% | 201-400 | 77 | 23% |
| **Age** | **#** | **%** | 401-800 | 24 | 7% |
| below 19 | 27 | 8% | 801-1000 | 10 | 3% |
| 20-30 | 77 | 23% | over 1000 | 3 | 1% |
| 31-40 | 111 | 33% | | | |
| 41-50 | 114 | 34% | | | |
| over 50 | 7 | 2% | | | |

**Table 2. Demographics of the Respondent Sample**

## Data Analysis

To evaluate the proposed research model, partial least squares (PLS) structural equation modeling (SEM) method was employed by following the guidelines of Benitez et al. (2017). As per Henseler and Sarstedt (2013), PLS-SEM method is applied to analyze complex research models. For testing the measurement and structural models, the ADANCO 2.0.1 software was used with 4999 bootstrap resamples (Dijkstra and Henseler 2015). The sub-constructs of the digital perfectionism a-priori research model were estimated by using the mode B regression weights (Dijkstra 2010).

## Content Validity

We posit that the measures of the sub-constructs possess substantial content validity as all of them were derived through well-established literature from psychology (Frost et al. 1990). Yet, its departure from its traditional form to 'digital' context meant that the items must be re-established for content validity (Nunnally 1967). By following the guidelines of McKenzie et al. (1999) the content validity was established. First, when deriving the initial draft of the instrument, all relevant literature was analyzed (Lynn 1986). Then, the survey instrument was shared with a panel of experts to review and assess the content of the instrument (American Educational Research Association 2002). We ensured that the panel of experts possess the necessary experience and qualifications regarding the subject and the context. The panel evaluated and critiqued the measures for each construct and commented on how well each measure assess the sub-constructs. Then, we conducted a pilot test of the survey instrument using 33 respondents. We conducted a quantitative assessment as per Lawshe (1975) and determined the CVR for each measure. As per the assessment, the minimum CVR value of 0.775 was observed (at a statistical significance of $p < 0.05$).

## Confirmatory Composite Analysis

To test the overall fit of the proposed digital perfectionism research model, we conducted confirmatory composite analysis by following the guidelines Henseler et al. (2014) and Benitez et al. (2017). The objective of conducting this analysis was to assess the formative sub-constructs of the proposed model and to identify any model misspecifications (Henseler et al. 2014). As per Henseler et al. (2014) by comparing the empirical correlation matrix with the model-implied correlation matrix through the examination of the standardized root mean squared residual (SRMR), unweighted least squares (ULS) discrepancy (dULS) and geodesic discrepancy (dG), the appropriateness of the composite model was assessed. Through the assessment of these measures our objective was to assess the difference between the observed correlation matrix and the model-implied correlation matrix (Benitez et al. 2016; Hu and Bentler 1999). Further, the measure of SRMR assist us in evaluating the average extent of the difference between the observed and predicted correlations as an absolute measure of (model) fit criterion. The SRMR of the digital perfectionism a-priori model was 0.019 – well below the recommended threshold of less than 0.080 – at the 0.05 alpha level (Dijkstra and Henseler 2015; Hu and Bentler 1999), with dULS =0.166 and dG = 0.017 indicating that we can confirm with a probability of 5% that the measurement structure of our proposed model is correct. As such, we can progress to assess the specific properties of our composite constructs.

## Construct Validity

The constructs of the digital perfectionism model demonstrated acceptable convergent and discriminant validity, with the AVE for all four constructs demonstrating values above 0.5 (Cenfetelli and Bassellier 2009; Fornell and Larcker 1981). Since the AVE of each construct in the model was higher than the variance shared between the construct and the other constructs in the model, it was evident that there were strong and satisfactory values for discriminant validity. In Table 3 the results of the AVE analysis are demonstrated.

| Table 3. Construct Correlation Matrix | | | | |
|---|---|---|---|---|
| | 1 | 2 | 3 | 4 |
| Mistakes uneasiness (1) | 0.821 | | | |
| Disproportionate standards (2) | 0.105 | 0.94 | | |
| Criticism antipathy (3) | 0.229 | 0.103 | 0.947 | |
| Excessive orderliness (4) | 0.201 | 0.11 | 0.108 | 0.837 |

**Table 3. Construct Correlation Matrix**

### Testing the Measurement Model

The measures of each construct were assessed for multi-collinearity amongst the measures by conducting variance inflation factors (VIF) (Diamantopoulos and Siguaw 2006; Henseler 2017). It was evident that the multi-collinearity is less likely in the data as the VIF from a regression of all the constructs in the model ranged between 1.63 and 2.59 (Diamantopoulos and Siguaw 2006). In Figure 2 the values of the measurement model test (significant at the level of 0.005 Alpha) (Henseler et al. 2016a) is depicted. The strong and significant values of the four constructs confirms the validity of deriving digital perfectionism construct. Overall, the constructs of the a-priori model explained 82% of the variance of digital perfectionism construct (the adjusted $R^2$ of 0.82). Considering model parsimony, while there may be other constructs explaining the remaining variance, the percentage of explanation herein is considered adequate. Moreover, the values in Figure 2 establishes the convergent and discriminant validity of the digital perfectionism model sub-constructs. As depicted in Figure 2, all the t-values of the outer model weights exceeded the one-sided cut-off of 1.645 levels significant at the 0.05 (*) alpha protection level. The one-sided test was deemed fitting for the analysis as in the model we only hypothesized a positive contribution of the formative components of digital perfectionism. Further, as per Diamantopoulos and Winklhofer (2001), each measure is explained by the linear regression of its latent construct and its measurement error. As such, the convergent validity of the constructs obeyed the rules defined by Benitez et al. (2017) and Benitez et al. (2017). Observing the relative contributions of each of the sub-constructs, it appears that 'criticism antipathy' makes the highest contribution to digital perfectionism, second by 'mistakes uneasiness,' followed by 'disproportionate standards' and lastly by 'excessive orderliness.'

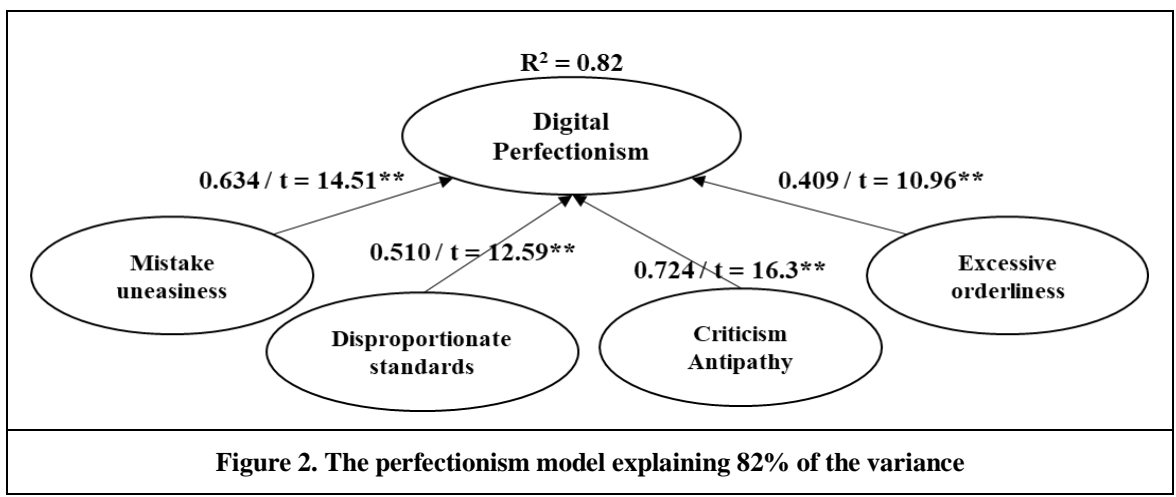

**Figure 2. The perfectionism model explaining 82% of the variance**

### Common Method Bias

Common method variance (CMV) was tested using (i) Harman's single factor test (Harman 1976), with largest factor accounted for 32% of the variance and rotated factor loading matrix showing that the items for each latent construct loaded on a single factor (while items for different constructs loaded on different factors); and (ii) confirmed that none of the significance levels of correlations among independent and dependent variables changed when we partial out common method bias using an unrelated "marker variable" (Lindell and Whitney 2001). As such, we do not believe that common method bias is a serious issue in this model.

### Structural Model

While the aforementioned tests demonstrate the existence of the digital perfectionism construct, the acid test determines how a newly developed construct behaves when it is introduced to a nomological network (Agarwal and Lucas Jr 2005). As per Cronbach and Meehl (1955), the nomological test is an important aspect of assessing the validity of the newly developed construct. The guidelines proposed by Cronbach and Meehl (1955) outline five steps in conducting construct validity using nomological network test. They are:

(i) the nomological net should at least have two constructs, (ii) the constructs should be theoretically or tautologically related, (iii) determining correspondence rule, (iv) determining empirical measurement of all constructs in the nomological network and (v) determining the empirical linkage through hypothesis development. Herein, we tested the theoretically purported relationship between digital perfectionism and *anxiety*[2], where past research has established that 'anxiety' arises through perfectionism (Hewitt and Flett 1991). In a simple hypothesis, one might conceive that 'perfectionism behavior leads to anxiety.' However, it is not a foregone conclusion that *digital* perfectionism would always lead to anxiety – given that any positive deviances would have positive effects in one's life. The path coefficient between digital perfectionism and anxiety was 0.7070 and the $R^2$ of the exogenous variable was 0.39 (39%). Not only did the results of the nomological network testing evidence the existence of a strong, positive and significant relationship between digital perfectionism and anxiety as hypothesized, they further evidenced the validity of both constructs (Diamantopoulos and Winklhofer 2001; Edwards and Bagozzi 2000)[3]. The impact of digital perfectionism seems to have a similar negative effect on one's level of anxiety. Past studies on perfectionism disorder (Grisham et al. 2010; Hamachek 1978; e.g., Henry and Crawford 2005; Purdon et al. 1999; Shaw et al. 2015) measured using the FMPS instrument, indicated the effect on anxiety ranges between $R^2$ 0.209 and 0.309. As such, the results established in this study, where digital perfectionism has a positive significant effect on anxiety, explaining 39% of its variance is also a worthy of a finding.

## Post-hoc Multi-group Analyses

Having validated the digital perfectionism construct and having demonstrated the nomological relationship between digital perfectionism and anxiety, we now explore possible conditions that are perhaps unique and exclusive to digital perfectionism. Employing the demographic details presented in Table 2, several categorical variables were derived for multi-group analyses, seeking better insights on digital perfectionism and its possible unique associations in the extended digital context.

These analyses are driven to explore answers for discussions that arise from the literature on factors such as: (i) number of associates in digital social circles, (ii) age, (iii) gender and (iv) whether the digital engagement is personal or professional. The four variables are also deemed important in the IS literature (e.g., Prensky 2001a; Prensky 2001b; Vodanovich et al. 2010). We performed four post-hoc multi-group analyses to explore whether there are statistically significant differences between several comparative groups identified through the sample (Sarstedt et al. 2017). Sarstedt et al. (2011, p. 198) note that "conceptually, the comparison of group-specific effects entails the consideration of a categorical moderator variable which, in line with Baron and Kenny (1986, p. 1174)." Therefore, it is hypothesized that the categorical variables derived below affects the direction and/or strength of the relation between digital perfectionism and anxiety (Henseler et al. 2009; Klesel et al. 2019).

The multi-group analysis commences with the MICOM (measurement invariance of composite models) procedure, which assesses the invariance[4] (Henseler et al. 2016b). Henseler et al. (2016b) recommend the use of MICOM, employing a three-step approach to analyze (i) configural invariance, (ii) compositional invariance, and (iii) the equality of composite mean values and variances. The results confirm the three types of invariance, which implies that measurement invariance holds and that a multi-group analysis is therefore possible[5] (Hair et al. 2018; Henseler et al. 2016b; Schlagel and Sarstedt 2016). After confirming the existence of invariance, the next step was to apply the multi-group analysis, comparing the explained variance for each group using appropriate categorical variables. In this case, (i) number of associates in digital social networks, (ii) personal or professional nature of engagement, (iii) age and (iv) gender as were used as categorical variables, using appropriate subsamples (e.g., male vs. female). The last step of the multi-group analysis was to analyze the differences between coefficients in the different paths. If these differences are significant[6], the categorical variable has a moderating effect.

---

[2] The nine reflective measures based on the study of anxiety loaded into a single factor indicating their suitability as a measurement construct.

[3] This further evidences the construct validity of digital perfectionism, by 'identification through structural relations' (Jarvis et al. 2003).

[4] Hult et al. (2008, p. 1028) note that "*failure to establish data equivalence is a potential source of measurement error (i.e., discrepancies of what is intended to be measured and what is actually measured), which accentuates the precision of estimators, reduces the power of statistical tests of hypotheses, and provides misleading results.*"

[5] Conceptually, measurement invariance expresses the idea that the measurement properties of X in relation to the target latent trait $W_i$ are the same across sub-samples. Meaning that the moderating variable has no effect on the loadings of the elements in the model.

### The Number of Associates in Digital Social Networks

First, we explored the effect of the emerging digital social networks of the individual. In traditional FMPS instrument, Frost et al. (1990) argue that an individual is less likely to receive substantial pressure from anyone other than *one's parents*. However, research on the influence on social media like Facebook, Twitter and LinkedIn have demonstrated that much of the influence nowadays come from external social circles (Alarifi et al. 2015; Chua and Chang 2016; Palekar and Sedera 2018; Sedera et al. 2017a). Moreover, as per Pempek et al. (2009) and consistent with Table 1, social networks have furthered comparison behaviors. Furthermore, the sheer magnitude of social interactions means that individuals are fraught with 'more' voices to be listened to (Lim 2016).

To test the effect of the number of associates in digital social networks, two polarized sub-samples were created using the top and the bottom 35 percentiles of the sample. Next, the two samples ($n_1$ of 129 respondents with less than 75 associates in social circles and $n_2$ of 117 respondents with over 199 associates) were assessed using the multi-group analyses described earlier. As per Table 4, when there are more associates (or friends) in digital social network, there is a higher likelihood of digital perfectionism leading to anxiety among individuals. This idea conforms with Sedera et al. (2017a), where social influence of associates in the digital social networks leading to dissatisfaction. Results in Table 4 demonstrate that there are significant differences between two sub-samples based on the number of associates in the digital social networks.

| Table 4. Multi-group analysis of sub-samples based on the number of associates | | | | | | |
|---|---|---|---|---|---|---|
| | $G_{1\ n > 75} = 129$ | $G_{2\ n < 199} = 117$ | $G_1$ vs. $G_2$ | | | |
| | $\beta_{n > 75}$ | $\beta_{n < 199}$ | $(\beta_{n > 75} - \beta_{n < 199})$ | t-value | Sig | p-value |
| Digital Perfectionism to anxiety | 0.504 | 0.628 | 0.124 | 5.163 | S | 0.001 |

**Table 4. Multi-group analysis of sub-samples based on the number of associates**

### Professional vs. Personal Engagement

We next investigate whether there is any significant difference in digital perfectionism based on one's type of engagements (i.e. personal vs professional engagement). Here, we employed the dichotomous scale to the question "My engagements with digital content is mainly professional / mainly personal." On one side, studies highlight that professional environments, regardless of the nature of engagement, can be demanding and can make an individual more anxious (Galluch et al. 2015). To the contrary, one could argue that, personal engagements in the digital spaces could be as demanding and stressful as the professional engagements (Padoa et al. 2018). For example, Padoa et al. (2018) described how individuals who recently became mothers become more anxious to promote their best selves on Facebook. As such, in order to test whether the personal vs. professional engagement increase the likelihood of digital perfectionism, two sub-samples were derived using demographic details in Table 2 ($n_{personal} = 128$ and $n_{professional} = 208$).

Table 5 depicts the multi-group analysis results for the sub-samples based on the nature of engagement (i.e. personal vs. professional). First, it illustrates that both sub-samples have significant and large path coefficients for the relationship between digital perfectionism and anxiety. More specifically, the PLS-multi-group analysis demonstrates no significant differences between the sub-samples, evidencing that individuals likely to experience anxiety through digital perfectionism regardless of the nature of engagement. This finding too conforms with the prior literature and anecdotal commentaries that despite the type of digital engagement, individuals attempt to promote their perfect self, thereby exerting additional pressure on themselves.

---

[6] If the p-value is smaller than 0.05 or larger than 0.95 for the difference of group-specific path coefficients (see Henseler et al., 2009).

| Table 5. Multi-group analysis based on the nature of digital engagement | | | | | | |
|---|---|---|---|---|---|---|
| | $G_{1\ pro}$=208 | $G_{2\ per}$ = 128 | $G_1$ vs. $G_2$ | | | |
| | $\beta_{pro}$ | $\beta_{per}$ | $(\beta_{pro} - \beta_{per})$ | t-value | Sig | p-value |
| Digital Perfectionism to anxiety | 0.483 | 0.556 | 0.073 | 0.684 | NS | 0.542 |

**Table 5. Multi-group analysis based on the nature of digital engagement**

## Age of the Respondent

Next, we investigated whether there are substantial differences in digital perfectionism leading to anxiety based on the respondent's age. Stoeber and Stoeber (2009) identify that the age is irrelevant for perfectionism leading to anxious behavior. However, studies (e.g., Vodanovich et al. 2010) argue that 'millennials' are more likely to have a positive associations with digital content (Prensky 2001a; Prensky 2001b), than their older counterparts. When considering the social media use of college students, Pempek et al. (2009) highlight that mostly the young adults tend to compare themselves with their friends on social media. While investigating the perfectionism behavior, Hewitt et al. (2017) identify that teenagers show a greater tendency toward perfectionism than their older peers. Curran and Hill (2019) highlight that one of the reasons for such tendency is that social media platforms' (i.e., Facebook, Instagram and Snapchat) capabilities to share digital content that curates the perfect version of oneself and lifestyle with others. Especially, individuals believe their social context (i.e. social media friends and followers) are extremely demanding, harsh and they must display perfect pictures of themselves to secure approval (Curran and Hill 2019). While such behaviors are common in social media platforms and evident that it is rising at an alarming rate, less attention has been received in prior literature.

To investigate this, two sub-samples were derived based on those who are considered millennials and Generation-Y (n millennials = 215 Gen-Y = 121)[7]. The multi-group analysis reported in Table 6 shows that both sub-samples demonstrated large, significant path coefficients between digital perfectionism and anxiety. Interestingly, the PLS-multi-group comparison revealed significant differences between these two age groups in the relationship between digital perfectionism and anxiety. As such, this finding too conforms with prior literature where they highlight the tendency among young adults feeling anxious due to perfectionism.

| Table 6. Multi-group analysis based on the age groups | | | | | | |
|---|---|---|---|---|---|---|
| | $G_{1\ Mille}$=215 | $G_{2\ GenY}$ = 121 | $G_1$ vs. $G_2$ | | | |
| | $\beta_{Mille}$ | $\beta_{GenY}$ | $(\beta_{Mille} - \beta_{GenY})$ | t-value | Sig | p-value |
| Digital Perfectionism to anxiety | 0.615 | 0.413 | 0.202 | 7.741 | S | 0.001 |

**Table 6. Multi-group analysis based on the age groups**

## Gender of the Respondent

Finally, we explored the gender differences of the sample for digital perfectionism leading to anxiety relationship. The two sub-samples derived using the demographic data yielded two samples (n male = 148 and n female = 188). As reported in Table 7, we found statistically significant differences between males and females on the relationship between digital perfectionism and anxiety. However, prior studies find that gender did not make a difference in traditional perfectionism behavior (Pacht 1984).

---

[7] Millennials are those who are born after 1980 and Gen-Y are those born years ranging from the early-to-mid 1960s to 1980. Given the data collection was completed in 2017, the maximum age of a millennial would be 39 years and the maximum age of a Gen-Y in the sample would be 53 years.

| Table 7. Post-hoc multi-group analysis based on the gender | | | | | | |
|---|---|---|---|---|---|---|
| $G_1$ Female =208 | | $G_2$ Male = 128 | $G_1$ vs. $G_2$ | | | |
| | $\beta$ Female | $\beta$ Male | ($\beta$ Female - $\beta$ Male) | t-value | Sig | p-value |
| Digital Perfectionism to anxiety | 0.583 | 0.413 | 0.170 | 5.591 | S | 0.001 |

**Table 7. Post-hoc multi-group analysis based on the gender**

## Conclusion

The objective of this paper was to investigate the notion of digital perfectionism – whether individuals manifest characteristics of perfectionism when engaging with digital content – by developing measures to assess its existence. This paper also investigated whether there is an association between digital perfectionism and anxiety among individuals. Perfectionism, in its traditional form, has been recognized as an independent diagnostic entity in the DSM-III (American Psychiatric Association 2013). As such, it has been hypothesized to play a major role in a wide variety of psychopathologies (Frost et al. 1990). In this paper, we urged the need for the IS community to seriously investigate the effects of such perfectionism behavior in relation to engaging with digital content.

Commencing with the conceptual discussion of Table 1 and the corresponding empirical evidences presented in the study, we argue that digital perfectionism behavior is strongly encouraged, endorsed and sought by individuals in the modern society of digitalization, both in personal and professional engagements. Given the polarized interactions between the four entities, the tolerance for mistakes becomes very low in relation to digital content. Such lack of tolerance is exacerbated due to accessibility and improvements in digital tools. Further, when dissemination of digital content on digital social networking sites, the social influence and the impact on the individuals are high compared to physical spaces where there are only a limited participants in the physical social circles. On the other hand, perfectionist behavior is facilitated through digital tools, software and equipment. For example, software that auto-corrects and beautifies photographs provide an unrealistic 'perfect image' of one-self as well as about one's peers. Our study highlighted some unfortunate circumstances of suicide and extreme depression when such perfectionist behavior is present. This emerging phenomenon is termed as 'digital perfectionism' in this research, retorting its roots in traditional perfectionism behavior established in DSM-III. Our study is motivated by the widespread anecdotal commentary that the diagnostic conditions of perfectionism seem to exist with most individuals in relation to digital content.

To investigate this phenomenon, we developed an a-priori model for digital perfectionism using four well-established sub-constructs based on nearly two-decades of work on perfectionism (Cox et al. 2002; Frost et al. 1990; Hollander 1965). The a-priori model, with the four sub-constructs of mistake uneasiness, disproportionate standards, criticism antipathy and excessive orderliness, was then tested using a sample of 336 respondents. We employed the PLS-SEM software ADANCO to test the a-priori model, following the guidelines of Benitez et al. (2018). The analyses, which was completed in to assess both measurement and structural models, demonstrated that the four sub-constructs adequately measure digital perfectionism. Next, the effect of digital perfectionism was observed using the individual's level of anxiety, where anxiety has been recognized as an important consequence of perfectionism (Cox et al. 2002; Frost and Marten 1990; Grisham et al. 2005; Pacht 1984; Purdon et al. 1999). The analyses revealed that digital perfectionism, like that of traditional perfectionism disorder, could cause higher levels of anxiety.

## Research Implications

In this paper, we have emphasized the importance of understanding digital perfectionism and the conceptualization of digital perfectionism using four sub-constructs as a formative composite construct. The items to measure each of the four sub-constructs were carefully constructed from an extensive literature review. Extending the validity of these sub-constructs (and its measures) to the paradigm of digital perfectionism, is a substantial contribution of this research. While theoretically based in DSM-III and its related studies, the digital perfectionism is not direct adaptation of the traditional perfectionism model.

Instead, our research logically contested the application of the sub-constructs, making important considerations related to IS. For example, the considerations of Table 1 on the creator, content, audience and their interactions with the tools are reflected in the conceptual design, survey instrument and the empirical findings of the study. Especially, the post-hoc analyses furthered our understanding of the considerations and their influence on digital perfectionism.

The final model of four sub-constructs explained 82% of the variance of digital perfectionism provides a good starting point for a cumulative tradition of research that has both parsimony and completeness. By demonstrating a strong nomological relationship between the independent variable (i.e. digital perfectionism) with anxiety as the dependent variable provided further confidence that the model is valid as a composite formative model. Further, this study becomes the first empirical study to assess the association between digital perfectionism and anxiety. Our treatment of digital perfectionism as a composite formative model provided extensive attention to the completeness, mutual exclusivity and necessity of the dimensions and measures of the model. Overall, the model statistics (e.g., variance explained, CVR, VIF and AVE) evidenced a strong model, with adequate attention to model parsimony. At the same time, the model evidenced significant differences in views on theoretically informed relationships of digital perfectionism.

In addition, the study also highlighted interesting results related to digital perfectionism and anxiety using post-hoc multi-group analyses. Through the analysis, it was evident that higher number of associates in the digital social networks will increase the anxiety due to digital perfectionism behavior. This finding further extends the understanding of social influence (Sedera and Lokuge 2018; Sedera et al. 2017a) and social impact on individuals in digital social networks (Palekar et al. 2015; Sedera et al. 2017b). Further, it was also evident that female users compared to male users are under pressure to display a perfect self in digital social networks and when engaging with digital content. Similarly, young adults (millennials) are comparing themselves with peers and attempt to portray a perfect self to others than Gen-Y users.

When there are ample opportunities to pursue perfectionist behavior, raising the awareness of this emerging phenomenon of digital perfectionism is a salient contribution of this research. When there are clear warning signs of digital perfectionism in the popular press (and evident in our daily lives), we were surprised with the scant attention that this topic has received in any discipline. While we are far from being able to claim that digital perfectionism as an illness, we believe we have commenced a vital discussion that would lead to a cumulative tradition of research. Unlike traditional perfectionism, if clinically established, digital perfectionism would have widespread repercussions due to the sheer number of people engaged with digital content. The growth of digitization initiatives world-wide (e.g., social media), digital mimicking of our daily routines and device-induced perfectionism through smart phones and digital cameras increase the chance of one's inclination to seek and encourage perfectionism behavior unknowingly. Raising an awareness of digital perfectionism would allow researchers of medical science, psychology, social science and management to contribute to a better knowledge of digital perfectionism.

## Practical Implications

This study makes several practical contributions. *First*, this study provides early empirical evidence of a potential psychologically damaging condition known as digital perfectionism. The validated sub-constructs and the items can be used to make an early diagnosis of the symptoms of digital perfectionism to gauge one's own status of digital perfectionism. *Second*, the validated model could lead to the development of management frameworks that can be derived to prevent and control digital perfectionism. As stated above, the authors did not find any evidence-based discussion on the dangers of digital perfectionism in any discipline. For example, when there is growing anecdotal evidence of suicide, depression and anxiety in relation to digital perfectionism behavior, some preliminary guidelines on how to deal with social criticism, avoiding disproportionate standards and not overly worrying about the mistakes can be beneficial. Therefore, the clinicians of psychology and social workers can pay proactive attention to this emerging issue.

## Limitations and Future Work

There are several limitations of this study. First, this study may run into a problem of defining digital perfectionism in this way as this may not distinguish neurotic perfectionistic people from those who are highly competent and successful. In the extreme, one might argue that setting of and striving for disproportionate standards is certainly not pathological. Similarly, some would argue that it reflects a positive outlook on life. While a healthy pursuit of excellence from those who take a genuine interest in striving to meet disproportionate standards is admirable, digital interactions, tools, equipment, software and the increasing number of friends in our digital social networks seem to provide an environment that allows inflated disproportionate standards beyond the reach of reasonable efforts, with people who strain compulsively and unremittingly toward near impossible goals. To conclusively address this deep phenomenon of interest, further testing through large and representative samples are required. We also encourage the use of tools like Amazon's Mechanical Turk for insight gathering exercises. The application of such tools will enhance the generalizability and validity of future studies and will provide more opportunities to explore more heterogeneity analyses. We recognize the weaknesses associated with a cross-sectional approach followed in this study, which is often criticized for its temporality. While a snapshot of digital perfectionism was precisely sought, we strongly encourage a longitudinal study to investigate how the validated sub-constructs and measures emerge and evolve.

Yoo, Y. 2010. "Computing in Everyday Life: A Call for Research on Experiential Computing," *MIS Quarterly* (34:2), pp 213-231.

## Appendix A – The Survey Instrument

| Construct | Measures |
|---|---|
| **Mistake uneasiness** (Adapted fromFrost et al. 1990; Stöber 1998; Stumpf and Parker 2000) | I should be upset if I make a mistake in my digital content. If someone creates digital content better than I, then I feel like I failed the whole task. I hate being less than the best at creating my digital content. People will probably think less of me if I make a mistake in my digital content. If I do not do as well as other people in creating digital content, it means I am an inferior individual. If I do not do well all the time in creating digital content, people will not respect me. The fewer mistakes I make in my digital content; the more people will like me. |
| **Disproportionate standards** (Adapted fromFrost et al. 1990; Stöber 1998; Stumpf and Parker 2000) | If I do not set the highest standards for myself in creating digital content, I am likely to end up as a second-rate person. It is important to me that I be thoroughly competent in creating my digital content. I set higher goals than most people in creating my digital content. I am very good at focusing my efforts in producing my digital content. I have extremely high goals in producing my digital content. Other people seem to accept lower standards in producing digital content from themselves than I do. I expect higher performance in producing digital content than most people. |
| **Criticism antipathy** (Adapted fromFrost et al. 1990; Stöber 1998; Stumpf and Parker 2000) | My friends, family and colleagues set very high standards for me when producing digital content. My friends, family and colleagues wanted me to be the best at everything when producing digital content. Only outstanding performance when producing digital content is good enough to my friends and family. My friends, family and colleagues have expected excellence from me when producing digital content. My friends, family and colleagues have always had higher expectations for my digital content. |
| **Excessive orderliness** (Adapted) | Organization of digital content is very important to me. Neatness in digital content is very important to me. |
| **Anxiety** (Lovibond and Lovibond 1995) | I find hard to wind down I find myself getting upset by quite trivial things I find myself getting agitated I tend to over-react to situations I find that I am very irritable I feel that I am rather touchy or impatient I am intolerant of anything that keeps me from getting on with what I was doing I find myself getting impatient when I am delayed in any way I find it difficult to tolerate interruptions to what I am doing |
| **Global measures** (Developed) | Overall, the digital content that I create are less prone to errors. Overall, I don't make mistakes in the digital content outputs. Overall, I maintain a good standard in all my digital content. Overall, I am confident about the quality of the digital content I produced. |